\documentclass[11pt]{article}
\usepackage{color} 
\usepackage{amsmath}
\usepackage{amssymb}
\usepackage{bm}
\usepackage[dvipdfm]{graphicx} 

\newcommand{\bea}{\begin{eqnarray}}
\newcommand{\eea}{\end{eqnarray}}
\newcommand{\be}{\begin{equation}}
\newcommand{\ee}{\end{equation}}

 \makeatletter  
\def\alt{\mathrel{\mathpalette\gl@align<}}
\def\agt{\mathrel{\mathpalette\gl@align>}}
\def\gl@align#1#2{ \lower.6ex\vbox{\baselineskip\z@skip\lineskip\z@
\ialign{ $\m@th#1\hfil##\hfil$\crcr#2\crcr\sim\crcr }} } \makeatother

\begin{document}
\begin{flushright}
HUPD-1315,
OIQP-14-6
\end{flushright}

%
\begin{center}

\vspace{1.5cm}
\baselineskip 20pt 
{\Large\bf 
Thermal property in Brownian motion of a particle 
\\coupled to vacuum fluctuations
}
\vspace{1.5cm}

{
\large
Naritaka Oshita${}^1$, Kazuhiro Yamamoto${}^{1,2}$, and Sen Zhang${}^3$
}
 \vspace{.5cm}

{\baselineskip 20pt \it
$^{1}$Department of Physical Science, Hiroshima University,\\ 
         Higashi-Hiroshima 739-8526, Japan \\
$^{2}$Hiroshima Astrophysical Science Center, Hiroshima University,\\
         Higashi-Hiroshima 739-8526, Japan \\
$^{3}$Okayama Institute for Quantum Physics, \\
Kyoyama 1-9-1, Kita-ku, Okayama 700-0015, Japan
}

\baselineskip 18pt
\vspace{2cm} 
\end{center}
\abstract{
We investigate Brownian motions of a particle coupled to vacuum 
fluctuations of a quantum field. 
The Unruh effect predicts that an observer in an accelerated motion sees 
the Minkowski vacuum as thermally excited. This addresses the problem of
whether a thermal property appears in a perturbative random motion 
of a particle in an accelerated motion due to the coupling or not. 
We revisit this problem by solving the equation of motion of a particle
coupled to vacuum fluctuations including the radiation reaction force. 
We compute a Fourier integral for the variance of the random velocity 
in a rigorous manner. Similarly, we consider a particle coupled to vacuum 
fluctuations in de Sitter spacetime motivated by the argument that an 
observer in de Sitter spacetime sees the Bunch-Davies vacuum as a 
thermally excited state with the Gibbons-Hawking temperature. 
Our investigation clarifies the condition that the energy equipartition 
relation arises in the Brownian motions of a particle. 
}
\thispagestyle{empty}
\newpage

\addtocounter{page}{-1}

\baselineskip 18pt

\section{Introduction}

Vacuum fluctuation is a key-concept of quantum physics in curved spacetime. 
One of the famous phenomena is the Hawking radiation, which predicts 
thermal-like radiation from black spacetime \cite{HawkingRadiation}.
From the equivalence principle, the Hawking radiation is closely related with 
the Unruh effect that predicts an observer in an accelerated motion sees the 
Minkowski vacuum as thermally excited \cite{Unruh}.
It is an interesting question whether the Unruh effect can be tested in a laboratory
or not. The radiation coming from the Unruh effect is called the Unruh radiation, 
though the existence of the Unruh radiation is under debate \cite{ChenTajima,ELI,IYZ}. 

Motivated by the arguments \cite{ChenTajima,ELI,IYZ}, 
we investigate Brownian motion of a particle due to the coupling to quantum vacuum fluctuations. 
This issue was first formulated in Ref.~\cite{IYZ}, in which the authors 
considered the system that consists of a particle and a quantum field, 
which are coupled to each other. 
They derived a Langevin-like equation for a particle taking the random force
from the quantum field fluctuations and the radiation reaction force into account. 
They applied the formulation to a particle in an uniformly accelerated motion, 
and found that the energy equipartition relation appears in the transverse 
fluctuations of a particle, which is perpendicular to the direction of 
the acceleration. 
However, the conclusion is based on a low energy approximation, and
the investigation is restricted to the transverse fluctuations in 
particle' motions because of the limitation of the approximation. 
Therefore, the longitudinal fluctuations have not been investigated so far.

In the present paper, complementary to the previous work~\cite{IYZ}, we focus our investigation
on the energy equipartition relation in random motions of a particle coupled to vacuum fluctuations
of a quantum field. 
We consider the system that consists of a particle and a quantum field, which are coupled to each other, in curved spacetime. We present basic formulas for a particle in random motions as a 
generalization of the previous work~\cite{IYZ}. As applications, 
we consider a particle in de Sitter spacetime and a particle 
in an uniformly accelerated motion, which is well described by 
the Rindler spacetime coordinate. 
We compute the variance of the random velocity in a coincidence limit of 
the two point function of the random velocity, which we evaluate by counting 
all the poles in the Fourier integral, which is obtained by solving the 
equation of motion of a particle. 
Our results clarify the difference between the transverse and the longitudinal 
fluctuations of a particle in an accelerated motion. 
The energy equipartition relation appears for the transverse fluctuations, 
but it does not appear for the longitudinal fluctuations when 
a particle is uniformly accelerated.

A particle coupled to vacuum fluctuations in de Sitter spacetime is also considered,
motivated by the prediction that an observer in de Sitter spacetime sees the Bunch-Davies 
vacuum as a thermally excited state with the Gibbons-Hawking temperature. 
For a particle coupled to the Bunch-Davies vacuum fluctuations in de Sitter spacetime, 
we find a similar structure of poles in the Fourier integral for the variance of 
the random velocity
as those of the longitudinal fluctuations of a particle in an accelerated motion. 
Our rigorous method shows that the energy equipartition relation does not appear 
for $m\gg H$, where $m$ is the mass of the particle and $H$ is a Hubble constant 
specifying the de Sitter expansion rate. The energy equipartition relation only 
appears for $m\ll H$. 

This paper is organized as follows: In section 2, we first present basis formulas, 
which describe the equation of motion of a particle coupled to vacuum fluctuations 
of a quantum field, as a generalization of the previous work \cite{IYZ} to 
those in curved spacetime.
We derive the Langevin-like equation for a particle taking
random forces due to the coupling and the radiation reaction 
force into account. In section 3, we consider a particle 
coupled to the Bunch-Davies vacuum fluctuations in de Sitter spacetime. 
In section 4, we consider a particle coupled to the Minkowski vacuum fluctuations 
in an accelerated motion. The transverse and the longitudinal random motions are
investigated separately. Section 5 is devoted to summary and conclusions. 
We follow the metric convention $(-,+,+,+)$.

\section{Basic formulas}
We consider the system consisting of a particle and a scalar field in 
curved spacetime, whose action is given by
\def\dotz{{\dot z}}
\begin{equation}
 S=S_{0}(z)+S_0(\phi)+S_{\rm int}(z,\phi),
\label{action2}
\end{equation}
where $S_0(z)$ and $S_0(\phi)$ are the action for the fee particle and the 
field conformally coupled to the curvature, 
\begin{eqnarray}
 &&S_{0}(z)=-m\int d\tau \sqrt{-g_{\mu\nu} \dotz^\mu \dot z^\nu},
\label{1-1}
\\
 &&S_0(\phi) =\int {d^4x} \sqrt{-g}\   \frac{1}{2} 
     \left\{ -g^{\mu\nu} \partial_\mu\phi  \partial_\nu\phi 
                  -\xi R\phi^2 \right\} ,
\label{1-2}
\end{eqnarray}
and $S_{\rm int}(z,\phi)$ describes the interaction, 
\begin{eqnarray}
 &&S_{\rm int}(z,\phi) 
  =  e\int d\tau {d^4x} \sqrt{-g_{\mu\nu}(x) \dotz^\mu \dot z^\nu} \phi(x) \delta^{4}\left(x-z(\tau)\right)
= e\int d\tau \sqrt{-g_{\mu\nu}(z(\tau)) \dotz^\mu \dot z^\nu} \phi(z(\tau)). 
\nonumber 
\\
\label{1-3}
\end{eqnarray}
Note that $x^\mu=z^\mu(\tau)$ denotes the trajectory of the particle, which 
obeys,
\begin{eqnarray}
 m{D \dotz^\mu\over D\tau}=e\left({D \dotz^\mu\over D\tau} \phi
 +\dotz^\mu \dotz^\alpha{\partial \phi \over \partial x^\alpha}
 +g^{\mu\alpha}{\partial \phi \over \partial x^\alpha}\right)\bigg|_{x=z(\tau)},
\label{emp}
\end{eqnarray}
where we used the notation $D/D\tau=\dot z^\mu\nabla_\mu$. 
Equation of motion for the scalar field is,
\begin{eqnarray}
  &&\left( -\nabla^\mu \nabla_\mu +\xi R \right) \phi(x) 
   =\frac{e}{\sqrt{-g}} \int d\tau \sqrt{-g_{\mu\nu}\dotz^\mu \dot z^\nu} 
 \delta^4 (x-z(\tau)).
\label{fieldeq}
\end{eqnarray}
The field equation has the solution, $\phi=\phi_{\rm h}+\phi_{\rm inh}$, where 
$\phi_{\rm h}$ and $\phi_{\rm inh}$ are the homogeneous and inhomogeneous 
solutions, respectively. The inhomogeneous solution is written as 
\begin{eqnarray}
\phi_{\rm inh}(x)&=&e \int d^4x' G_R(x,x')\int d\tau' \sqrt{-g_{\mu\nu}\dotz^\mu \dot z^\nu}  
\delta^4 (x'-z(\tau'))
\nonumber\\
&=&e \int^\tau d\tau' G_R(x,z(\tau')),
\end{eqnarray}
where  $G_R(x,y)$ denotes the retarded Green function, which satisfies 
$
{\sqrt{-g}}\left( -\nabla^\mu \nabla_\mu+\xi R \right) G_R(x,y) 
   =\delta^4(x-y).
$

The terms from the inhomogeneous solution in Eq. (\ref{emp})
give rise to a radiation reaction force for the equation of 
motion of the particle \cite{DeWitt,Hobbs1,LH,GHL,IYZ}. 
We consider the conformally flat spacetime, in which it is known that the tail term 
in the radiation reaction force disappears \cite{Hobbs}.
The terms from the homogeneous solution $\phi_{\rm h}$ in Eq. (\ref{emp}) give 
rise to random forces, and we have the Langevin-like equation of motion 
\begin{eqnarray}
 &&m{D \dotz^\mu\over D\tau}={e^2 \over 12\pi} \left({D^2 \dotz^\mu\over D\tau^2}-\dotz^\mu
\Bigl({D \dotz\over D\tau}\Bigr)^2\right)
-{e^2\over 24\pi}(g^{\mu\nu}+\dot z^\mu\dot z^\nu)R_{\nu\alpha}\dot z^\alpha
\nonumber\\
&&~~~~~~~~~~~~~+e
\left({D \dotz^\mu\over D\tau} \phi_{\rm h}
 +\dotz^\mu \dotz^\alpha{\partial\phi_{\rm h} \over \partial x^\alpha}
 +g^{\mu\alpha}{\partial \phi_{\rm h} \over \partial x^\alpha}\right)\bigg|_{x=z(\tau)},
\label{stochasticeq}
\end{eqnarray}
where the mass of the particle is redefined.
This equation is a generalization of that derived in \cite{IYZ}, in which a particle 
in an uniformly accelerated motion in the Minkowski spacetime is considered.
Using this generalized equation, we first consider 
the particle in de Sitter spacetime in section 3. 

\section{A particle in de Sitter spacetime}
We consider the particle in de Sitter spacetime, 
whose line element is written as 
\begin{eqnarray}
  ds^2=-dt^2+a^2(t)\delta_{ij} dx^i dx^j,
\end{eqnarray}
where $a(t)=e^{Ht}$ is the scale factor. The equation of motion  
is derived directly from Eq.~(\ref{stochasticeq}). 
We write the trajectory of the particle $x^\mu=z^\mu(\tau)=(t(z),z^i(\tau))$, 
and derive the linearized equation of motion of $z^i(\tau)$, 
assuming that the particle moves around the origin of the spatial 
coordinate. Then we have
\begin{eqnarray}
&&m (\dot v^i+H v^i)={e^2\over 12\pi}(\ddot v^i+H \dot v^i)
+{e\over a}{\partial \phi_{\rm h} \over \partial x^i}\Bigr|_{x=z(\tau)},
\end{eqnarray}
where we defined $v^i(\tau)=a(\tau) \dot z^i(\tau)$.
This equation is solved in the Fourier space as
\begin{eqnarray}
\widetilde V^i(\omega)=e{\widetilde \varphi_i(\omega)\over (\omega -iH)(e^2 \omega/12\pi+im)},
\end{eqnarray}
where we defined the Fourier expansion, 
\begin{eqnarray}
&& v^i(\tau)=\int {d\omega \over 2\pi} \widetilde V^i(\omega) e^{-i\omega\tau},
\\
&& {1\over a(\tau)}{\partial \phi\over \partial x^i}\Bigr|_{x=z(\tau)}
=\int {d\omega \over 2\pi} \widetilde \varphi_i(\omega) e^{-i\omega\tau}.
\end{eqnarray}
As we consider the scalar field conformally coupled to the curvature
in the Bunch Davies vacuum, we have \cite{BD,Murata},
\begin{eqnarray}
\biggl \langle 0 \biggr|
{1\over a(t_x)}{\partial\phi_{\rm h}(x)\over \partial x^i}
{1\over a(t_y)}{\partial\phi_{\rm h}(y)\over \partial x^j}\biggr| 0\biggr\rangle 
\biggr|_{x=z(\tau),y=z(\tau')}
={H^4\over 32\pi^2}{\delta_{ij}\over \left(\sinh H(\tau-\tau')/2-i\epsilon\right)^4}, 
\end{eqnarray}
and
\begin{eqnarray}
&&\langle{v}^i(\tau){v}^j(\tau')\rangle_S=\delta_{ij}
 {6\over e^2}\int_{-\infty}^\infty d\omega{\omega\over \omega^2+(12\pi m/e^2)^2}\coth(\pi \omega/H) e^{i\omega(\tau'-\tau)},
\end{eqnarray}
where '$\langle\cdots\rangle_S$' means the symmetrization with respect to $\tau$ and $\tau'$.
The poles of the integrand are $\omega=\pm i12\pi m/e^2$, and $\pm inH$ with $n=1,2,\cdots$.  
The integration can be performed exactly. 
Then, we have
\begin{eqnarray}
\langle{v}^i(\tau){v}^j(\tau')\rangle_S&=&{6\delta_{ij}\over e^2}\left[
e^{-H\sigma^{-1}\delta\tau}\pi\cot \pi \sigma^{-1}+\sum_{n=1}^\infty\left({1\over n+\sigma^{-1}}
+{1\over n-\sigma^{-1}}\right)e^{-nH\delta\tau}
\right]
\nonumber
\\
&=&\delta_{ij} {H\over 2\pi m}\Xi[H,\sigma^{-1};\delta\tau],
\end{eqnarray}
where we defined $\sigma=e^2H/12\pi m$, and  $\delta\tau\equiv|\tau'-\tau|$.
In the second line, we introduced the function defined by
\begin{eqnarray}
\Xi[H,\zeta;\delta\tau]=\zeta\left\{e^{-H \zeta \delta\tau}\pi \cot(\pi \zeta)+\sum_{n=1}^\infty
\left({1\over n+\zeta}+{1\over n-\zeta}\right)e^{-n H\delta\tau}
 \right\}.
\label{defxi}
\end{eqnarray}
With the use of the mathematical formulas, 
\begin{eqnarray}
&&\sum_{n=0}^\infty{e^{-n a \delta\tau}\over n+\zeta}={}_2F_1(1,\zeta,1+\zeta; e^{-a\delta\tau})
{1\over \zeta}
\\
&&{}_2F_1(a,b,a+b;z)={\Gamma(a+b)\over\Gamma(a)\Gamma(b)}\sum_{n=0}^\infty
{(a)_n(b)_n\over (n!)^2}
\biggl[2\psi(n+1)
\nonumber\\
&&~~~~~~~~~~~~~~~~~~~~~~~
-\psi(a+n)-\psi(b+n)
-\ln(1-z)
\biggr](1-z)^n,
\\
&&\psi(n+z)=\sum_{m=0}^{n-1}{1\over m+z}+\psi(z),
\end{eqnarray}
where ${}_2F_1(a,b,c;z)$ is the hypergeometric function, $\psi(z)$ is the poly-gamma function and $\Gamma(a)$ is the gamma function, and the symbol $(a)_n$ is defined as 
$(a)_n=a(a+1)(a+2)\cdots(a+n-1)$, $\Xi[H,\zeta;\delta\tau]$ 
can be rewritten as
\begin{eqnarray}
&&\Xi[H,\zeta;\delta\tau]=\zeta\biggl\{(e^{-H\zeta\delta\tau}+e^{H\zeta\delta\tau})
\Bigl(-\gamma-H\delta\tau-\psi(\zeta)-\ln(1-e^{-H\delta\tau})\Bigr)
-{e^{-H\zeta\delta\tau}\over \zeta}
\nonumber
\\
&&\hspace{1.5cm}
-{\partial \over \partial c}
{}_2F_1(1,\zeta,c;1-e^{-H\delta\tau}) \bigg|_{c=1}
-{\partial \over \partial c}
{}_2F_1(1,-\zeta,c;1-e^{-H\delta\tau}) \bigg|_{c=1}\biggr\},
\end{eqnarray}
where $\gamma$ is the Euler constant.
In the limit of small $\delta\tau$, this function has the asymptotic 
formula,
\begin{eqnarray}
\Xi[H,\zeta,\delta\tau]\simeq-1+2\zeta(-\ln (H\delta\tau)-\gamma-\psi(\zeta))
+{\cal O}(\delta\tau).
\label{approxi}
\end{eqnarray}
Noting the definition $\sigma^{-1}=12\pi m/He^2$ and 
using the asymptotic formula of the poly-Gamma function,  
\begin{eqnarray}
\psi(\zeta)=
\left\{
\begin{array}{ll}
\displaystyle{\ln \zeta-{1\over 2\zeta}-{1\over 12\zeta^2}}, & {\rm for}~~\zeta\gg1,
\\
\displaystyle{-{1\over \zeta}-\gamma}, & {\rm for}~~\zeta\ll1,
\end{array}
\right.
\label{appropg}
\end{eqnarray}
we finally have 
\begin{eqnarray}
\langle{v}^i(\tau){v}^j(\tau')\rangle_S=\delta_{ij}\times
\left\{
\begin{array}{ll}
\displaystyle{-{12\over e^2}\Bigl(\ln {12\pi m\delta\tau\over e^2} +\gamma\Bigr) +{\cal O}\Bigl((H/m)^2\Bigr)} & ~{\rm for}~~H \ll 12\pi m/e^2, 
\\
\displaystyle{{H\over 2\pi m}-{12\over e^2}\ln (H\delta\tau) +{\cal O}(m/H)} &~{\rm for}~~ H \gg 12\pi m/e^2.
\end{array}
\right.
\label{VVdeS}
\end{eqnarray}
The results show that the variance of velocity has a logarithmic divergence in the limit 
that $\delta\tau$ approaches zero. 
However, since the coefficient of the divergence terms are independent of $H$, 
these terms don't contribute to the thermal nature. Indeed the term $-{12\over e^2}
\left(\ln {12\pi m\delta \tau \over e^2} +\gamma\right)$
remains in the limit of $H \rightarrow 0$, 
where the de Sitter spacetime reduces to the Minkowski spacetime.
As will described in the next section, we also find that the same term 
remains for the case of particle in an uniformly accelerated motion in the limit of the 
acceleration constant approaches zero. So one may
understand that this divergence comes from the short-distance 
motion of the particle, originated from our formulation based on point particle.  
The divergence coming from the short-distance motion of the particle 
will be removed by taking a finite size effect of the particle into account.
Therefore, this suggests that $\delta\tau$ cannot be taken to be
zero, and it is natural to introduce a finite value cutoff~\cite{JTHsiang}.

Eq.~(\ref{VVdeS}) means that the energy equipartition relation does not 
appear for the case $H\ll 12\pi m/e^2$, where one cannot find a thermal 
property in the random motion. 
On the other hand, the energy equipartition relation looks to appear when 
$H\gg 12\pi m/e^2$ as the Gibbons-Hawking temperature is $H/2\pi$.
It is well known that the scalar field in de Sitter spacetime has
a stochastic property \cite{starobinsky,sasaki,Rigopoulos}. 
Especially, the stochastic inflation approach indicates
that the massless scalar field coarse-grained over the horizon size $H^{-1}$ 
follows an equation of motion with a white noise term. 
The results (\ref{VVdeS}) in the case $H\gg 12\pi m/e^2$ would be closely 
related to this fact. 
However, we should also note that the amplitude of the velocity is larger 
than unity, even though the energy equipartition relation appears.
This inconsistency would come from the fact that $H\gg m$ means that the
particle's Compton wavelength is much longer than the horizon size.
This would mean the inconsistency of our description of the point particle in this 
regime. 

\section{Accelerated particle}
Now let us consider the particle in an accelerated motion with an uniform 
acceleration $a$. 
Equations of motion around an uniformly accelerated motion
is derived in Ref.~\cite{IYZ}, which are summarized in the following.
We consider the transverse and the longitudinal random motions separately.
\subsection{transverse fluctuations} 
For the transverse fluctuations perpendicular to the direction of the 
uniform acceleration, the equation of motion is 
\begin{eqnarray}
&&m\dot v^i={e^2\over 12\pi}(\ddot v^i-a^2 v^i)+{e}{\partial \phi_{\rm h} \over \partial x^i}\Bigr|_{x=z(\tau)},
\label{teq}
\end{eqnarray}
which leads to \cite{IYZ}
\begin{eqnarray}
\langle v^{i} (\tau) v^{j} (\tau ') \rangle_{S} = \delta_{ij} \frac{e^2}{6} 
\int_{-\infty}^\infty  d\omega \frac{\omega (\omega^2 + a^2)}{ \bigl( {e^2 (\omega^2 + a^2)}\bigr)^2 +(12\pi m \omega)^2} 
\coth(\pi\omega/a)
e^{i \omega\delta\tau}.
\label{vvtransvers}
\end{eqnarray}
The integrand has the poles at $\omega= \pm ia\Omega_+,\pm ia \Omega_-$, and $\pm ina$ with $n=2,3,\cdots$, 
where $\Omega_\pm$ are defined as
\begin{eqnarray}
&&\Omega_\pm=\sqrt{1+{1\over 2\sigma^2}\pm\sqrt{\left(1+{1\over 2\sigma^2}\right)^2-1}}
=\frac{1}{2\sigma} \left(\pm 1+\sqrt{1+4 \sigma^2} \right),
\end{eqnarray}
and here $\sigma$ is defined as $\sigma={e^2 a/12\pi m}$. 
%
%
This integration can be exactly performed, and we have
\begin{eqnarray}
&&\hspace{-1cm}
\langle{ v}^i(\tau){ v}^j(\tau')\rangle_S=\delta_{ij}
{a\over 2\pi m} {1\over \sqrt{1+4\sigma^2}}
\biggl[\Omega_+e^{-a\Omega_+\delta\tau}\pi \cot\pi\Omega_+
+\Omega_+\sum_{n=2}^\infty\left({1\over n+\Omega_+}+{1\over n-\Omega_+}\right)e^{-na\delta\tau}
\nonumber
\\&&~~~~~~~~~~~~~~~~~~~+\Omega_-e^{-a\Omega_-\delta\tau}\pi \cot\pi\Omega_-
+\Omega_-\sum_{n=2}^\infty\left({1\over n+\Omega_-}+{1\over n-\Omega_-}\right)e^{-na\delta\tau}
\biggr],
\end{eqnarray}
where we used 
\begin{eqnarray}
{1-\Omega_-^2\over \Omega_+^2-\Omega_-^2}=
{1\over 2}\left(1-{1\over \sqrt{1+4\sigma^2}}\right)
,~~
{1-\Omega_+^2\over \Omega_-^2-\Omega_+^2}=
{1\over 2}\left(1+{1\over \sqrt{1+4\sigma^2}}\right).
\end{eqnarray}
Then, using the function $\Xi[a,\Omega_\pm;\delta\tau]$ 
defined by Eq.~(\ref{defxi}), we have 
\begin{eqnarray}
&&\langle{ v}^i(\tau){ v}^j(\tau')\rangle_S=
{\delta_{ij}}{a\over 2\pi m} {1\over \sqrt{1+4\sigma^2}}\Bigl(
\Xi[a,\Omega_+;\delta\tau]+\Xi[a,\Omega_-;\delta\tau]
\Bigr).
\end{eqnarray}
Using the asymptotic formula (\ref{approxi}) with replacing $H$ 
with $a$, we have 
\begin{eqnarray}
&&\langle{\delta v}^i(\tau){\delta v}^j(\tau')\rangle_S
=\delta_{ij}\frac{a}{2 \pi m} \frac{2}{\sqrt{1+4\sigma^2}} \Bigl\{ -1 -\Omega_+ \psi ( \Omega_+ ) -\Omega_- \psi ( \Omega_- ) 
\nonumber
\\
&&~~~~~~~~~~~~~~~~~~~~~~~~~~~~~~~~~~~~~~~~~~~~~~~~~~
+ ( \Omega_+  + \Omega_- ) ( -\log (a\delta\tau) - \gamma ) \Bigr\}+{\cal O}(\delta\tau).
\label{jojo}
\end{eqnarray}
In the limit of small $\delta\tau$, with the approximate formula (\ref{appropg}), 
expression (\ref{jojo}) reduces to
\begin{eqnarray}
\langle{v}^i(\tau){v}^j(\tau')\rangle_S=
\delta_{ij}\times
\left\{
\begin{array}{ll}
\displaystyle{{a\over 2\pi m}-{12\over e^2}\Bigl(\ln {12\pi m\delta\tau\over e^2}+\gamma\Bigr) 
+{\cal O}\Bigl((a/m)^2\Bigr)} &~ {\rm for}~ a \ll 12\pi m/e^2, 
\\
\displaystyle{-{12\over e^2}\Bigl(\ln ({a\delta\tau}) +{1\over2}\Bigr) +{\cal O}(m/a)} &~
{\rm for}~a \gg 12\pi m/e^2.
\end{array}
\right.
\end{eqnarray}
For the case $a \ll 12\pi m/e^2$, the equipartition relation 
appears as the Unruh temperature is $a/2\pi$, which 
was first discovered in Ref.~\cite{IYZ}. On the other hand,
it does not for the case $a \gg 12\pi m/e^2$, however, 
this case perhaps should not be considered because  
the Unruh temperature is much larger than the particle mass.

\subsection{longitudinal fluctuations}
Now we consider the longitudinal fluctuations, whose equation of motion is 
given by (see appendix in Ref.~\cite{IYZ}), 
\begin{equation}
m ( \delta\ddot\xi -a^2 \delta \xi) = \frac{e^2}{12 \pi} \Bigl( \delta \dddot\xi -
a^2 \delta \dot \xi \Bigr) +e \vartheta \phi_{h}\big|_{x=z(\tau)},
\label{16}
\end{equation}
where $\vartheta \equiv \Bigl( a+{\partial}/{\partial \xi} \Bigr)$, 
$\delta\xi(\tau)$ specifies the position of the fluctuating particle 
in the longitudinal direction, and $\delta\xi=0$ is the trajectory
of the particle without fluctuations.
This equation can be derived from (\ref{stochasticeq}) with an external force 
using the Rindler coordinate, 
\begin{eqnarray}
  ds^2=e^{2a\xi}(-d\eta^2+d\xi^2)+dx_1^2+dx_2^2.
\end{eqnarray}
The use of the Fourier expansion leads to the following formula for the variance of the 
random velocity
\begin{eqnarray}
  &&\langle\delta \dot\xi(\tau)\delta\dot\xi(\tau')\rangle_S
={6 \over e^2} \int_{-\infty}^\infty  d \omega \frac{\omega^3}{a^2+\omega^2} \frac{e^{i \omega \delta \tau}}
{(12\pi m/e^2)^2 + \omega^2} \coth(\pi \omega/a).
\end{eqnarray}
The poles of the integrand are $\pm i\sigma^{-1}$, $\pm ina$ with $n=1, 2,\cdots$. 
The poles at $\pm ia$ are the second order.
The structure of the pole is similar to that of a particle in 
de Sitter spacetime. 
The integration can be exactly evaluated as
\begin{eqnarray}
&&\langle\delta \dot\xi(\tau)\delta\dot\xi(\tau')\rangle_S
={6\over e^2}\biggl[e^{-a\delta\tau/\sigma}\pi \cot \pi\sigma^{-1}+{\sigma^4-5\sigma^2-2a\delta\tau\sigma^4+2a\delta\tau\sigma^2\over 2(1-\sigma^2)^2}e^{-a\delta\tau}
\nonumber\\
&&~~~~~~~~~~~~~~~~~~~~~
+\sum_{n=2}^\infty{2n^3e^{-na\delta\tau}\over (\sigma^{-2}-n^2)(1-n^2)}\biggr].
\end{eqnarray}
By using the function $\Xi[a,\zeta;\delta\tau]$
defined by Eq.~(\ref{defxi}) with replacing variable $H$ and $\zeta$ with 
$a$ and $\sigma^{-1}$, respectively, we may rewrite
\begin{eqnarray}
&&\langle\delta \dot\xi(\tau)\delta\dot\xi(\tau')\rangle_S
={6 \over e^2} \biggl[{\sigma\over 1-\sigma^2}\Xi[a,\sigma^{-1};\delta\tau]
+{2\sigma^2e^{-a\delta\tau}\over (1-\sigma^2)^2}
+{\sigma^4(1-5\sigma^{-2}-2(1-\sigma^{-2})a\delta\tau)\over 2(1-\sigma^{2})^2}e^{-a\delta\tau}
\nonumber
\\
&&~~~~~~~~~~~~~~~~~~~~~~~~~~
+{\sigma^2\over 1-\sigma^{2}}\Bigl((e^{a\delta\tau}+e^{-a\delta\tau})\ln(1-e^{-a\delta\tau})+1+
{1\over 2}e^{-a\delta\tau}\Bigr)\biggr], 
\end{eqnarray}
which reduces to
\begin{eqnarray}
&&\langle\delta \dot\xi(\tau)\delta\dot\xi(\tau')\rangle_S
={6 \over e^2} \biggl[-2\log(a\delta\tau)-{2\gamma+2\psi(\sigma^{-1})
+\sigma-\sigma^2\over 1-\sigma^2}\biggr]+{\cal O}(\delta\tau),
\end{eqnarray}
in the limit of small $\delta\tau$ with the use of Eq.~(\ref{approxi}). 
Furthermore, the use of the approximate formula (\ref{appropg}) leads to
\begin{eqnarray}
&&
\langle\delta \dot\xi(\tau)\delta\dot\xi(\tau')\rangle_S=
\left\{
\begin{array}{ll}
\displaystyle{-{12\over e^2}\Bigl(\ln {12\pi m\delta\tau\over e^2}+\gamma\Bigr) 
+{\cal O}\bigl((a/m)^2\bigr)} &~~~ {\rm for}~~a \ll 12\pi m/e^2,
\\
\displaystyle{-{12\over e^2}\Bigl(\log(a\delta\tau)+{1\over 2}\Bigr) 
+{\cal O}(m/a)} &~~~ {\rm for}~~a \gg 12\pi m/e^2.
\end{array}
\right.
\end{eqnarray}
Thus, the results indicate that the energy equipartition relation does not appear in
the longitudinal fluctuations, which is the contrast to the transverse fluctuations. 
This difference comes from the second term in the left hand side of equation of motion 
(\ref{16}), $-ma^2\delta\xi$. Such a term in a classical dynamics makes its motion unstable. 
This prevents the energy equipartition relation in the longitudinal direction. 
There is the blocking of the energy equipartition relation in the 
direction of the accelerated motion.

\section{Conclusions} 
In the present work, we investigated Brownian random motions of a particle
coupled to vacuum fluctuations. 
We consider a particle coupled vacuum fluctuations in de Sitter spacetime 
and a particle in an uniformly accelerated motion in the Minkowski spacetime. 
A detector coupled to vacuum fluctuations is excited as if 
it is exposed to the thermal bath with the Gibbons-Hawking temperature 
in de Sitter spacetime or with the Unruh temperature of an acceleration motion.
Then, we may expect that a thermal property may appear in the random motions 
of a particle in such situations. 
We investigate this problem by solving the equation of motion 
of a particle coupled to vacuum fluctuations including the radiation 
reaction force. 
We found that the energy equipartition relation may appear in the 
random motion of a particle. We also found that it does not always appear. 
For the particle in de Sitter spacetime, the energy equipartition relation
apparently appears when $H/m\gg1$, but it does not appear when $H/m\ll1$.
For the particle in the accelerated motion, the equipartition relation
appears in the transverse motion when $a/m\ll1$. 
We showed that the energy equipartition relation does not appear 
in the longitudinal fluctuations.
In the previous work \cite{IYZ}, the authors found that the energy 
equipartition relation appears in the transverse motion when 
$a/m\ll 1$, but the conclusion was derived by evaluating only the pole of the 
lowest energy scale in the Fourier integration.
In the present paper, our conclusions have been derived by evaluating 
all of the poles in the Fourier integration. 
The condition that the energy equipartition appears is that the lowest energy 
pole in the Fourier integral for the variance of velocity is smaller than the
Unruh temperature for the particle in the accelerated motion 
or the Gibbons-Hawking temperature for the particle in de Sitter spacetime. 
It is interesting to investigate whether the random motions representing 
a thermal property can be detected as a signature of radiation or not, 
which will be investigated in a future work. 

\section*{Acknowledgments} 
We would like to thank S.~Iso, H.~Kodama, M.~Sasaki, and T.~Tanaka
for useful conversation related with the topic of the present paper
at the early stage of this work. 
K.Y. acknowledges  useful discussions during the YITP Long-term 
workshop YITP-T-12-03 on "Gravity and Cosmology 2012".

\appendix

\end{document}